\documentclass[manuscript]{aastex}

\begin{document}

\title{Discovery of a Second Nova Eruption of V2487 Ophiuchi}
\author{Ashley Pagnotta, Bradley E. Schaefer, Limin Xiao, Andrew C. Collazzi}
\affil{Department of Physics and Astronomy, Louisiana State University, Baton Rouge, LA 70803}
\email{pagnotta@phys.lsu.edu}

\and

\author{Peter Kroll}
\affil{Sternwarte Sonneberg, D-96515 Sonneberg, Germany}

\begin{abstract}

A directed search for previously-undiscovered nova eruptions was conducted in the astronomical plate archives at Harvard College Observatory and Sonneberg Observatory.  We found that an eruption of V2487 Oph (Nova Oph 1998) occurred on 1900 June 20.  V2487 Oph was previously classified as a classical nova, which we identified as a probable recurrent nova based on its large expansion velocities and the presence of high excitation lines in the outburst spectrum.  The event was recorded on Harvard plate AM 505, at a B magnitude of $10.27\pm0.11$, which is near peak.  The outburst can only be seen on one plate, but the image has a characteristic ÔdumbbellÕ shape (caused by a double exposure) that is identical to the other star images on the plate, and thus is not a plate defect.  We conclude that this is in fact a previously-undiscovered nova outburst of V2487 Oph, confirming our prediction that it is a recurrent nova.  We also examine the discovery efficiency for eruptions of the system and conclude that a randomly-timed outburst has, on average, a 30\% chance of being discovered in the past century.  Using this, we deduce a recurrence time for V2487 Oph of approximately 18 years, which implies that the next eruption is expected around 2016.

\end{abstract}


\section{Introduction}

Recurrent novae (RNe) are cataclysmic variables which have undergone more than one nova eruption.  They consist of a white dwarf accreting material from a companion star until a runaway thermonuclear reaction is triggered on the surface of the white dwarf.  The short recurrence timescale separates the RNe from the classical novae (CNe), which likely recur with a timescale of up to $10^{5}$ years.  To have such a short recurrence time, all RNe systems must have both a high-mass white dwarf and a high accretion rate \citep{sekiguchi95, townsley09}.  The presence of a near-Chandrasekhar mass white dwarf that is quickly accreting more mass makes it likely that RNe will collapse as Type Ia supernovae, thus RNe are a probable solution to the Type Ia progenitor problem.  A central question is whether there are enough RNe in our galaxy to produce the observed Ia supernova rate.  To answer this, we need better information on the properties and discovery efficiency of RNe.

Many nova eruptions are missed ({\it c.f.} Section \ref{sec:stats}), leading to the likelihood of a number of RNe for which only one eruption has been detected.  This raises the question of just how many recurrent systems are currently mislabeled as classical.  The only efficient way to find new RNe is to examine the current list of CNe for systems that are actually recurrent.  (The alternative is to continuously monitor the entire sky for many decades, waiting for systems to erupt again.)  Many CNe systems have been proposed to be RNe, on many different grounds \citep{duerbeck86, harrison92, kato03, kato04, weight94, shears07}, but the only way to prove a system to be recurrent is to find a second nova event.  

Archival photographic plates provide coverage of the whole sky over a time period that extends back more than a century, and thus provide an effective way to search for previous outbursts of nova systems for which only one outburst has been observed.  Roughly three-quarters of the useful archival plates in existence are located either at the Harvard College Observatory in Cambridge, Massachusetts or at the Sonneberg Observatory in Sonneberg, Th\"{u}ringen, Germany.  Twenty of the thirty-five known RN eruptions are found on plates at Harvard and/or Sonneberg; in eleven cases these plates are the only record of the eruption.  Fourteen of the nineteen RN eruptions that occurred from 1890 to 1953 were actually discovered on plates.  Thus, in seeking to find RNe masquerading as CNe, a thorough examination of the Harvard and Sonneberg plates for previously-undiscovered eruptions is a logical step.

To make efficient use of time, it is best to select a subset of the CNe that are good candidates for RNe.  Previously, \citet{duerbeck86} noted that all RNe have fast declines and low eruption amplitudes.  Seven of the nine RNe have long orbital periods, and four of the nine have red giant companion stars.  These facts give us observational properties that can be used to select good RN candidates.  We have also identified two spectral characteristics displayed by the RNe during outburst.  From a search through the known CNe for these properties, we selected V2487 Oph (Nova Oph 1998) as our first target.  This paper reports on our test of this prediction: an exhaustive search of all relevant plates at Harvard and Sonneberg.  We discovered a prior eruption of V2487 Oph in 1900, confirming that the system is in fact recurrent.

\section{Prediction}

We considered a number of characteristics for each nova system, each of which has a range or particular value that indicates a possible RN.  An orbital period longer than 0.6 days is characteristic of most recurrent systems; in systems where there is no direct measurement of the orbital period, the observation of an infrared excess (pointing to a red giant companion) also implies a long period and is therefore of interest.  The \citet{duerbeck86} criterion is that RNe fall into a ``region void of classical novae" on a plot of eruption amplitude vs.  $t_{3}$, where  $t_{3}$ is the time it takes the brightness to fall three magnitudes from peak.  We have identified two new RN indicators that are easily observable in outburst spectra.  First, we note that {\it all} RNe display expansion velocities in excess of 2000 km s$^{-1}$, as measured by the full width at half maximum of the H$\alpha$ line, while most CNe have much lower velocities.  Additionally, we find that {\it all} RNe outburst spectra feature high-excitation lines such as [Fe X] (and higher) and He II.  Few of the CNe have He II or iron lines with ionization higher than [Fe II] during outburst.

Using this set of criteria, V2487 Oph rapidly came to our attention.  The nova was discovered by K. Takamizawa at magnitude 9.5 on 1998 June 15.561 \citep{nakano98}.  We have constructed a light curve from AAVSO observations, \citet{liller99}, \citet{hanzl98}, and our own CCD observations at quiescence in 2002 and 2003.  From this light curve we derive V$\mathrm{_{max}} = 9.5$, B$\mathrm{_{max}} = 10.1$, $t_{3} = 8$ days, and an amplitude of 8.2 mag.  This fast, low-amplitude nova easily satisfies the Duerbeck criterion.  From its outburst spectrum \citep{lynch00}, we see an expansion velocity of 10,000 km s$^{-1}$ and He II lines.  V2487 Oph is therefore a likely RN.

V2487 Oph has previously been suspected to be a recurrent nova.  \citet{hachisu02} describe the system as a ``strong candidate recurrent nova" based on their analysis of the light curve from the 1998 outburst.  They call particular attention to the rapid decline and the plateau phase of the outburst light curve, both of which are characteristic of fast RNe such as U Sco.  Based on the mass transfer rate in quiescence, they estimate a recurrence time of 40 years.  They also note that the system has a high-mass white dwarf accretor (M$_{WD} \approx 1.35\pm 0.01M_{\odot}$), as based on their model, which makes it a prime candidate for a Type Ia supernova progenitor in addition to being a probable RN.  \citet{rosenbush02} also names V2487 Oph as a potential recurrent nova, and places it in his CI Aql group, based on the shape of the outburst light curve.  \citet{hernanz02} also point to V2487 Oph as in interesting case, as it was apparently seen before the 1998 outburst in X-rays, as part of the ROSAT All-Sky Survey.

\section{Discovery}

Harvard College Observatory has approximately 500,000 archival plates, taken from the 1890s to the 1950s, and from the 1960s to 1980s.  There are two main types of plates: patrol plates and deep plates.  Patrol plates cover a large area of the sky, typically 20$^{\circ}$ square, and have a limiting magnitude $\sim$14.  Deep plates cover a smaller area of the sky, but go much deeper, often to around 18th magnitude.  Many of the Harvard plates for V2487 Oph were checked by one of us in 2004.  In the summer of 2008 we completed an exhaustive search of all plates at Harvard showing V2487 Oph.  The 1900 outburst of V2487 Oph was discovered at magnitude $\mathrm{B}=10.27\pm0.11$ on plate AM 505, which was taken on 1900 June 20 (JD 2415191.617).  AM 505 is part of the AM patrol plate series taken at Arequipa, Peru, is centered at right ascension 17$^{h}$ and declination -15$^{\circ}$, has a blue-sensitive emulsion, and has a limiting magnitude of $\mathrm{B} = 11.3$.  

All other plates housed at Harvard that could possibly show the 1900 outburst were examined.  There are no plates covering the position of V2487 Oph to a sufficient depth during the months surrounding the eruption to provide confirmation.  On plate B 25522 it can be seen that V2487 Oph was not up on 1900 June 2.  Plate I 25510 covers the area on 1900 June 30, but has a limiting magnitude of $\mathrm{B} = 11.3$.  By this time, approximately 10 days after peak, V2487 Oph had decreased in brightness to $\mathrm{B} \sim 13.6$ and thus there is no chance of seeing it on the plate.  During this time period, patrol images of a field were only taken approximately every two weeks, and deep plates were only taken for specific observing campaigns.  When combined with the short duration of the V2487 Oph outburst, it is therefore not surprising that no other plates show evidence of the outburst.  While there are no confirmation plates, there are none for which the lack of detection is inconsistent with an outburst on 1900 June 20.

\begin{figure}
\centering
\epsscale{.68}
\plotone{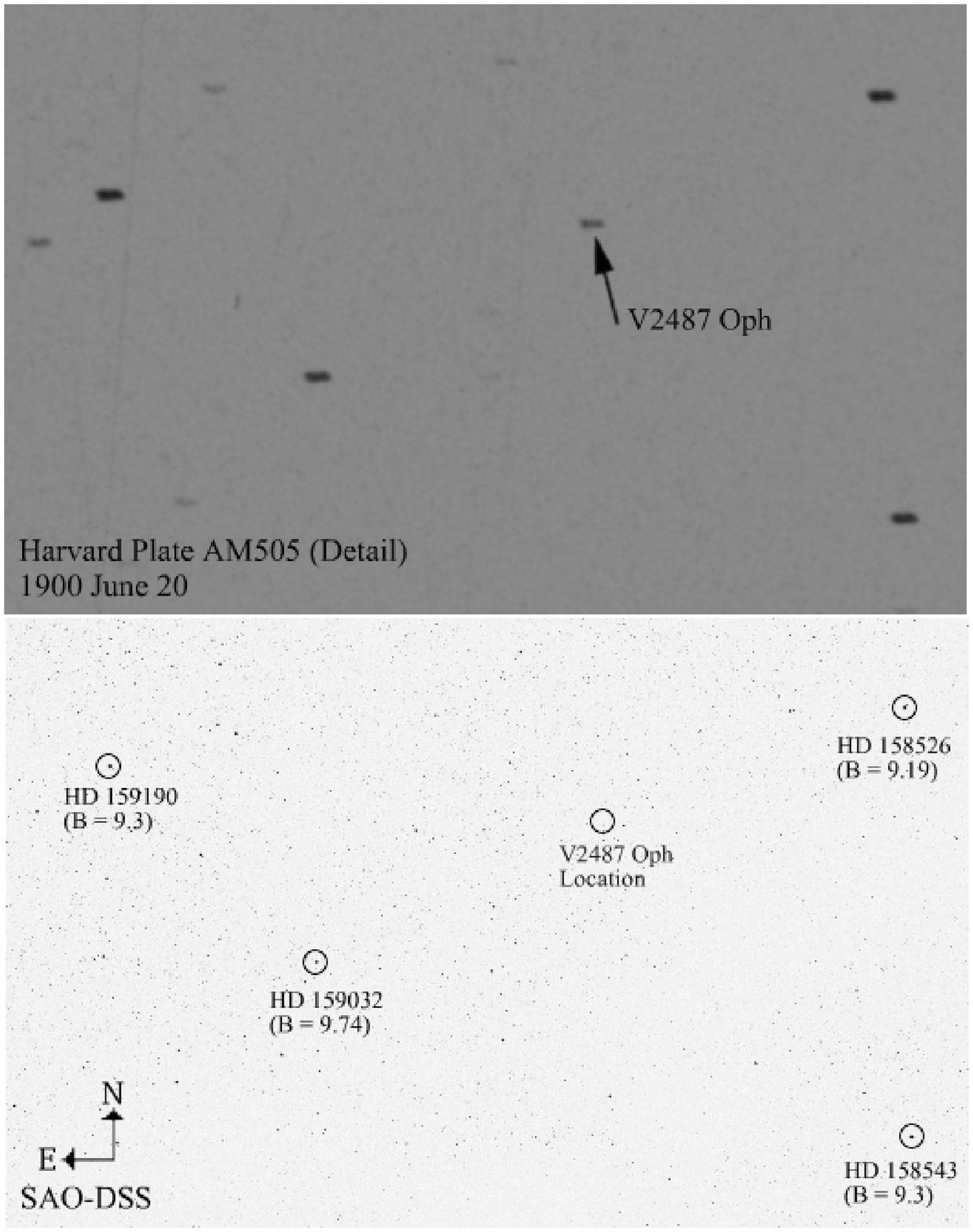}
\caption{{\it Top:} Scanned image of Harvard plate AM505.  V2487 Oph is marked with the arrow.  The field is approximately 60$'$ wide by 30$'$ tall; north is up, and east is to the left.  The 1.8$'$ image trail of V2487 Oph is identical to that of the neighboring stars, providing evidence that the image is not a plate defect, a near-telescope light source, or a solar system object.  The image is within 15$''$ (0.025mm on the plate) of the sky position of V2487 Oph, which is a small fraction of the length of the trail.  {\it Bottom:} SAO-DSS image of the same field.  The four most prominent stars from the plate are marked and labeled, as is the position of V2487 Oph.}
\label{plate}
\end{figure}

With only one plate showing the outburst, we must provide reasons for our confidence in this discovery.  A 1.8$'$ trailing in right ascension gives all of the star images on the plate a distinct ÒdumbbellÓ shape.  In Figure \ref{plate}, it can be seen that the object at the location of V2487 Oph has the same dumbbell shape, and therefore the light forming the image came from a fixed location in the sky.  This provides strong proof that the image is not any type of plate defect.  Trailing such as this is not unusual on the Harvard plates, though most plates show a more typical (circular) point spread function (PSF).  In this case, the imperfect tracking provides a unique PSF which, when combined with the fact that the image is in focus, indicates that the light originated beyond the Earth's atmosphere and then passed through it on its way to the telescope, a further argument against a plate defect or local light source.  The fact that the trail over the 61-minute exposure has the same length and orientation as the trails of nearby stars demonstrates that the point source cannot be an asteroid or other Near-Earth Object.  The Minor Planet Center confirms that there were no minor planets in this location at that time.  

Quantitative astrometry on the scanned plate places the star image in the correct location in relation to other stars in the area to within 2$''$ in declination and 15$''$ in right ascension.  For this plate series, 15$''$ corresponds to 0.025 mm on the plate.  The limiting magnitude of the plate precludes the possibility of the image being a background star, except in the case that the background star was undergoing a flare greater than 10 magnitudes, which is extremely improbable.  In all, despite having only one image, we have strong proof that the light forming the image was from a point source above the telescope, outside the Earth's atmosphere and our Solar System, and from the location of a specific nova suspected to have had prior eruptions.  Thus we conclude that we found a previously-undiscovered nova eruption of V2487 Oph.  Combined with the nova outburst observed in 1998 \citep{nakano98}, we can be confident that V2487 Oph is a recurrent nova.

In addition to the Harvard plates, we also checked all of the available plates in the archive at the Sonneberg Observatory.  Sonneberg has approximately 300,000 plates, the majority of which were taken in Germany, so the southern coverage is not as complete as at Harvard.  Despite this, there are still approximately 1,500 plates that cover V2487 Oph during the time span from 1926 to 1994.  While there was no possibility of finding a confirmation plate for the 1900 outburst, we continued the search in an attempt to find more undiscovered outbursts of V2487 Oph.  It is reasonable to expect that there should be other outbursts because of purely statistical reasons ({\it c.f.} the following section on Discovery Statistics), and because of the 40 year recurrence time estimated by \citet{hachisu02}.  We did not find any additional outbursts.

\section{Discovery Statistics}
\label{sec:stats}

Nova eruptions are missed for many reasons, such as the solar gap when the nova is too close to the sun to be observed, the lunar gap when nova searchers generally pause due to the brightness of the Moon, and the scarcity of nova searchers who observe to faint magnitudes over the whole sky.  (\citet{shafter02} shows that half of the novae peak fainter than 7th mag.)  Our detailed study of observing times and limits from the many nova hunters and archival plate collections around the world shows a discovery efficiency ranging from 0.6\% to 19\% over the last century, for undirected searches, depending on the peak magnitude and $t_{3}$ of the system \citep{schaefer09}.  We distinguish between directed and undirected searches because there is a much higher chance of catching a system in outburst during a directed search, {\it i.e.} when we monitor a known recurrent system or search through the archives because there is reason to believe the system has had multiple nova eruptions.  

Most novae, however, are discovered during undirected searches; they just happen to be observed, often during completely unrelated observations, or during general variable star searches.  \citet{shafter02} has presented a comparison of model and observed magnitude distributions which demonstrates that only $\sim$10\% of faint nova eruptions (with an apparent $\mathrm{m_{peak}} > 8$ mag) are discovered.  Our calculations based on the known properties of the RNe indicate that the ratio of RNe currently mislabeled as CNe to the known RNe is roughly ten to one; since we now know of ten RNe, there are $\sim$100 recurrent systems currently known as CNe \citep{schaefer09}.

What is the likelihood of discovering an eruption?  An operational definition of the discovery efficiency is the fraction of the days in a year on which the nova can peak and the eruption would be detected.  This will depend on the dates and limiting magnitudes of the observations, as well as the eruption light curve of the nova.  Let us here work out an equation for calculating the discovery efficiency ($\epsilon$).  The coverage can usually be reasonably idealized as having some nearly constant limiting magnitude, and the nova light curve will be brighter than that limit for some length of time, $t_{vis}$.  For example, the light curve of V2487 Oph is brighter than $\mathrm{B} = 11$ mag for 3 days, and brighter than $\mathrm{B} = 14$ mag for 13 days (Schaefer 2009).  If we had a plate series that covered the position of V2487 Oph to $\mathrm{B} = 14$ on every night of the year, then the discovery efficiency would be unity because all eruptions that year would be discovered.  

The reality, however, is that most sets of observations have a significant gap of duration $G_s$ due to the passage of the Sun close to the position of the nova.   An eruption during most of this time period will go undiscovered.  The length of the solar gap depends on the series of observations, the latitude of the observers, and the declination of the nova.  To give typical examples for other RNe, the Harvard plates have average solar gaps of 100 days for T Pyx, 150 days for U Sco, and 235 days for RS Oph. To take a specific example, suppose we had a plate series that went to $\mathrm{B} = 14$ for 183 consecutive days in a year, with the solar gap covering the other half of the year.  In such a case, a V2487 Oph eruption could be discovered if the nova peaked anytime during the 183 consecutive days {\it plus} anytime in the last 13 days of the solar gap, as the declining tail of the eruption would be seen on the first post-gap plate.  The eruption would be discovered if the peak occurred on any of 196 days in the year, for an efficiency of 196/365=0.54.  In general for this simple case, the efficiency would be $\epsilon = [365-(G_s - t_{vis})]/365$.  In this equation, the subtracted quantity ($G_s-t_{vis}$) must never be allowed to become negative.  The numerator should be the number of days in the year on which the nova could peak and the eruption be discovered.  To give an important example, the recurrent nova T Pyx has a solar gap of near zero for AAVSO observations ($G_s \approx 0$) and yet the nova is brighter than the typical AAVSO limiting magnitude of $\textrm{V}=14.0$ mag for a duration of 270 days ($t_{vis}=270$ days), so we take the $G_s-t_{vis}$ value to be zero.  With this, $\epsilon = 1.0$, and we realize that any eruption of T Pyx after 1967 would certainly have been discovered.

The observations which might discover an eruption are almost never conducted nightly outside the solar gap, and indeed most data series have substantial gaps even when the nova is well placed in the sky.  Moon interference is a common cause of these gaps: if the moon is bright and close to the system, the limiting magnitude of the observations is too bright to be useful.  Because of this, many observers do not work on nights with a bright moon, causing gaps in the observational record.  Short gaps in the record can arise for other reasons.  The most common cause is simply a lack of observations, for example during the early years at both HCO and Sonneberg when fewer plates were taken, so the ones that do exist are spaced farther apart.  For purposes of calculating $\epsilon$, the cause of the gap does not matter; they are all taken together and called lunar gaps.  In practice, these gaps vary from month to month and year to year, but typical gaps are 15-20 days for lunar gaps and 20-40 days when gaps are caused by sparse records.  

Ideally, every gap would be meticulously recorded, but in reality the recording of every plate time would provide no useful improvement in accuracy.  Instead, we have recorded all plate times for many sample years and use average observed gap durations ($G_l$).  The number of days on which the peak would be missed due to a lunar gap is $G_l-t_{vis}$, again with this number never being allowed to fall below zero.  For example, a fast nova with $t_{vis}=10$ days, when covered by a source with a 20 day lunar gap, could be observed, on average, if it erupted on 20 days in a given month --  the 10 days at the end of the lunar gap and the 10 days outside the lunar gap.  If the nova is slow enough so that the eruption would be visible on the plates for longer than the lunar gap, then the existence of the lunar gap does not impact the discovery efficiency. Multiple lunar gaps occur during the time outside the solar gap; on average there are $N_l$ lunar gaps per year.  With this, the number of possible nova peak days that would result in a missed eruption is $N_l(G_l-t_{vis})$.  The final general equation for the discovery efficiency is then
\begin{eqnarray}
\epsilon= \frac{365 - (G_s - t_{vis}) - N_l (G_l - t_{vis})}{365}      
\end{eqnarray}
where the terms in parentheses cannot go negative.  Patrol plates are taken regularly, at evenly-spaced intervals, providing regular gap structures and good coverage during the observable times, at a cadence that is usually fast enough to catch most nova events.  Deep plates are not taken with as much regularity as the patrols; the coverage depends strongly on the field and the time period, and therefore deep plates do not always contribute significantly to the overall discovery efficiency of a particular system.  The total number of plates enters into this calculation only to the extent that the average gap durations can be reduced when there are a lot of plates, but in practice, an increase in the number of observations does little to decrease the gap durations because the sun and bright moon still dominate.  We now have a general way of calculating the discovery efficiency, and this only depends on the length of time that the nova would be visible and on the gap structure of the data series.

Let us give a typical worked example for V2487 Oph.  For the year 1990, the dominant detection source is the plate archive at Sonneberg. The average solar gap for these plates is $G_s = 210$ days, and there are four lunar gaps averaging $G_l = 15$ days each. The limiting magnitude of the plates is typically $\textrm{B} = 12$ mag. Based on the 1998 eruption light curve, a V2487 Oph eruption would be visible for $t_{vis} = 6$ days total. The discovery efficiency is therefore
\begin{eqnarray*}
 \epsilon 
   = \frac{365 - (G_s - t_{vis}) - N_l (G_l - t_{vis})}{365}      
   = \frac{ 365-(210-6)-4\times(15-6)}{365}
   = 0.34
 \end{eqnarray*}
for V2487 Oph in 1990.  That is, had the eruption peaked on any of 34\% of the nights in 1990, the eruption would have been recorded on the Sonneberg plates and discovered by us.

The low amplitude and short $t_3$ of V2487 Oph are two of the reasons it was a strong candidate RN, but they also lead to a short amount of time during which the eruption is brighter than plate limits, and therefore a low discovery efficiency.  Because of these characteristics, it is possible for an outburst to happen completely during a solar or lunar gap, leaving no observable evidence on the plates.

\begin{figure}
\centering
\epsscale{1}
\plotone{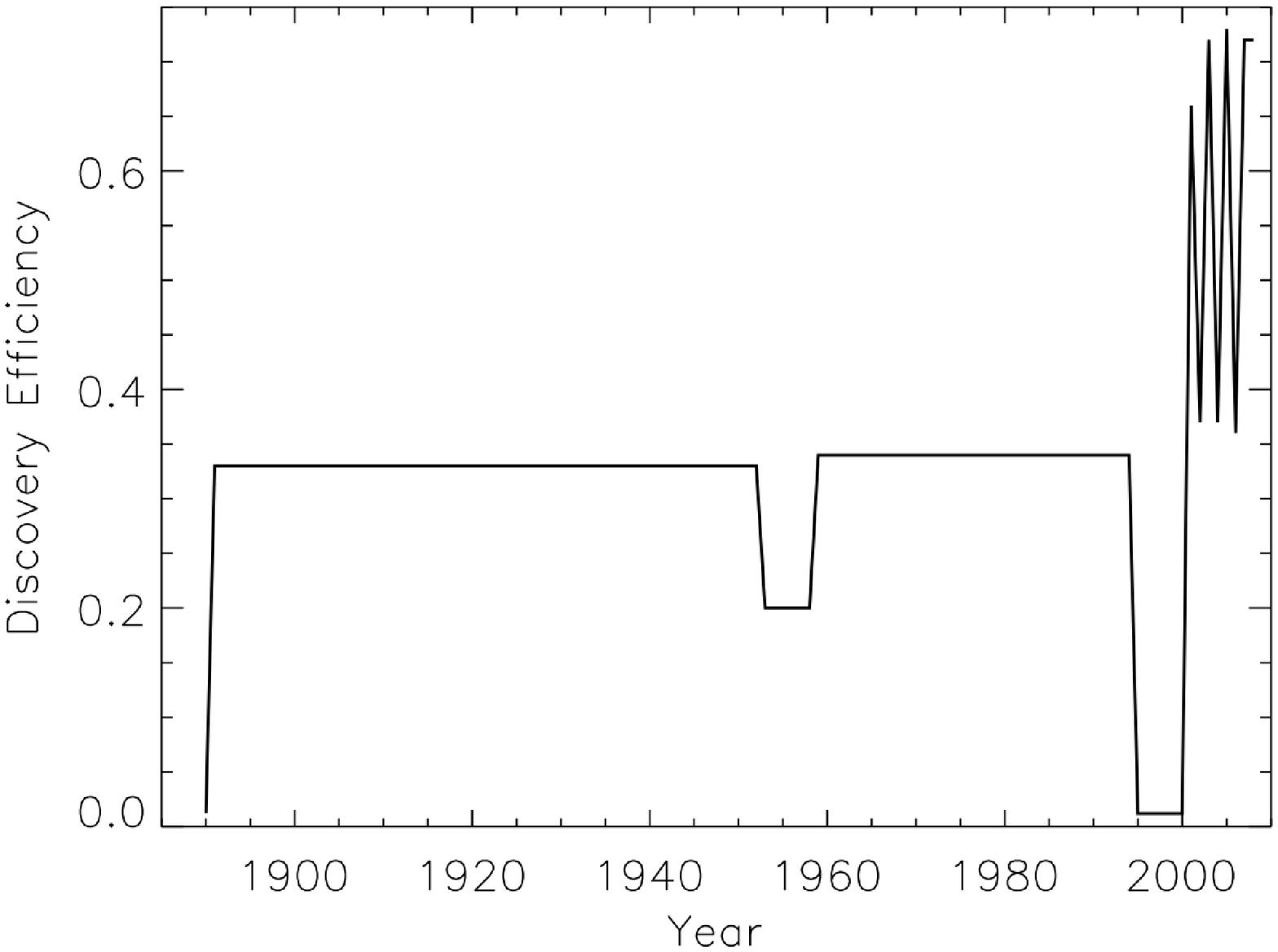}
\caption{This plot shows the yearly discovery efficiency for V2487 Oph, the percentage of days each year on which the nova can erupt and be detected.  The detection sources we consider are undirected photographic and visual searches, the Harvard plate collection, the Sonneberg plate collection, the All Sky Automated Survey (ASAS-3), and the observations of the American Association of Variable Star Observers.  The average discovery efficiency over the past 118 years is 30\%.  With two eruptions discovered, we estimate the total number of eruptions to be $2/0.30=6.7$, which implies that four or five eruptions have likely been missed.  This indicates an average eruption timescale of 18 years, and the possibility of the next eruption occurring as soon as 2016.}
\label{graph}
\end{figure}

The nova searches we considered are undirected photographic and visual searches by amateurs and professionals, the Harvard plate collection, the Sonneberg plate collection, the All Sky Automated Survey (ASAS-3) \citep{pojmanski02}, and the observations of the American Association of Variable Star Observers. For V2487 Oph, we have calculated the average detection rate on a year-by-year basis.  These results can be seen in Figure \ref{graph}.  From 1890 to 2008, the discovery efficiency of V2487 Oph ranges from 1\% to 73\%, and averages 30\%.  The years before the beginning of plate archive coverage are not considered, because V2487 Oph has such a low apparent magnitude that the probability of detection is negligible.

We can estimate the number of missed eruptions and the recurrence time scale in two ways.  First, we have used a Monte Carlo simulation to calculate the probability that a given recurrence time scale would produce exactly two discovered eruptions given the yearly probability of discovery from Figure \ref{graph}.  Supposing that the recurrence time scale is 98 years, then the average probability that a randomly-phased series of eruptions separated by 98 years will have exactly two discovered eruptions is 1.2\%.  Supposing that the recurrence time scale is $98/2=49$ years, then the mean probability that a randomly-phased series of eruptions separated by 49 years will have exactly two discovered eruptions is just 9.1\%.  Supposing that the recurrence time scale is 98/N years, then the average probability that a randomly phased series of eruptions will produce exactly two discovered eruptions will continue rising as N rises from 1 to 5 and then start falling.  For example, with $\textrm{N}=10$, we would have expected many more than two eruptions to have been discovered since 1890.  The probability of getting exactly two eruptions is $>30\%$ for $4 \le \textrm{N} \le 7$, while the probability is $<10\%$ for N values of 1, 2, and $>12$.  The most probable hypothesis is that $\textrm{N}\sim 5$.  The $\textrm{N}=5$ case corresponds to a recurrence time scale of 20 years and a total of six eruptions in the last 118 years, four of which were missed.  

Using the second method, we can directly estimate the most likely number of eruptions since 1890 by using the number of discovered eruptions and the 30\% average discovery efficiency.  Thus, the total number of eruptions should be 2/0.30, or 6.7 total eruptions in the past 118 years (four or five of which were missed).  The average recurrence time scale is then $118/6.7=18$ years.  Because of the large uncertainties in these methods, this is not inconsistent with the 40 year recurrence time predicted by \citet{hachisu02}.

\section{Implications} 

V2487 Oph is now the tenth known galactic recurrent nova.  Because of the importance of RN demographic information to the Type Ia supernova progenitor problem, and the small number of known RNe, it is important to obtain more information on V2487 Oph.  In particular, its orbital period, spectrum and spectral energy distribution at quiescence, and long- and short-term photometric variability should be investigated.  As a recurrent nova with a recurrence time on the order of 18 years, V2487 Oph could have another outburst as soon as 2016.  The monitoring of V2487 Oph should be increased, both by amateurs and professionals.  

Our discovery of this second eruption of V2487 Oph was a direct result of the test of our prediction that the system was in fact recurrent.  With this success, we have gained substantial confidence in the utility of identifying probable RNe based on high expansion velocities and high-excitation lines in outburst spectra, in addition to previously-published indicators.  It is highly likely that there are many more mislabeled RN systems for which only one eruption has been seen.  Identification of the best RN candidates, archival searches, and frequent monitoring of these candidates should be performed to discover more RNe and to obtain the best possible RN demographic information.

\acknowledgments

We greatly appreciate the help and hospitality of Alison Doane and Jaime Pepper at Harvard and the 4$\pi$ Systeme staff at Sonneberg, especially Klaus L${\rm \ddot o}$chel for his insights into the modern process of large-scale sky imaging on plates and films.  We also acknowledge the AAVSO staff and the observers worldwide who contribute to the AAVSO International Database.  Funding for this research was provided by NSF Grant AST-0708079.

\begin{thebibliography}{}

\bibitem[Duerbeck(1986)]{duerbeck86}
Duerbeck, H.W. 1986, in RS Ophiuchi (1985) and the Recurrent Nova Phenomenon, ed. M. F. Bode (Utrecht, The Netherlands: VNU Science Press), 99

\bibitem[Hachisu {\it et al.}(2002)]{hachisu02}
Hachisu, I., Kato, M., Kato, T., \& Matsumoto, K.  2002, in ASP Conf. Proc. Vol. 261, The Physics of Cataclysmic Variables and Related Objects, ed. B. T. G${\rm \ddot a}$nsicke, K. Beuermann, \& K. Reinsch (San Francisco: ASP), 629

\bibitem[Hanzl(1998)]{hanzl98}
Hanzl, D. 1998, IAU Circ. 6976

\bibitem[Harrison(1992)]{harrison92}
Harrison, T. E. 1992, \mnras, 259, 17

\bibitem[Hernanz \& Sala(2002)]{hernanz02}
Hernanz, M., \& Sala, G. 2002, Science, 298, 393

\bibitem[Kato \& Hachisu(2003)]{kato03}
Kato, M., \& Hachisu, I. 2003, \apj, 598, L107

\bibitem[Kato {\it et al.}(2004)]{kato04}
Kato, T., Yamaoka, H., \& Kiyota, S.  2004, \pasj, 56, 83

\bibitem[Liller \& Jones(1999)]{liller99}
Liller, W., \& Jones, A. 1999, IBVS 4774

\bibitem[Lynch {\it et al.}(2000)]{lynch00}
Lynch, D. K., Rudy, R. J., Mazuk, S., \& Puetter, R. C. 2000, \apj, 541, 791

\bibitem[Nakano(1998)]{nakano98}
Nakano, S. 1998, IAU Circ. 6941

\bibitem[Pojmanski(2002)]{pojmanski02}
Pojmanski, G. 2002, Acta Astronomica, 52, 397

\bibitem[Rosenbush(2002)]{rosenbush02}
Rosenbush, A. E.  2002, in AIP Conf. Proc. Vol. 637, Classical Nova Explosions: International Conference on Classical Nova Explosions, ed. M. Hernanz \& J. Jos\'e (Melville, NY: AIP), 294

\bibitem[Schaefer(2009)]{schaefer09}
Schaefer, B. E. 2009, \apjs, Submitted

\bibitem[Sekiguchi(1995)]{sekiguchi95}
Sekiguchi, K. 1995, Astrophysics \& Space Science, 230, 75

\bibitem[Shafter(2002)]{shafter02}
Shafter, A. W. 2002, in AIP Conf. Proc. Vol. 637, Classical Nova Explosions: International Conference on Classical Nova Explosions, ed. M. Hernanz \& J. Jos\'e (Melville, NY: AIP), 462

\bibitem[Shears \& Poyner(2007)]{shears07}
Shears, J., \& Poyner, G.  2007, JBAA, 117, 136

\bibitem[Townsley(2009)]{townsley09}
Townsley, D. M. 2009, in ASP Conf. Series Vol. 401, RS Ophiuchi (2006) and the Recurrent Nova Phenomenon, ed. A. Evans, M. F. Bode, T. J. O'Brien, \& M. J. Darnley (San Francisco: ASP), 131

\bibitem[Weight {\it et al.}(1994)]{weight94}
Weight, A., Evans, A., Naylor, T., Wood, J. H., \& Bode, M. F. 1994, \mnras, 266, 761

\end{thebibliography}
\end{document}